# Interlayer Fermi polarons of excited exciton states in quantizing magnetic fields

Huiying Cui[1,2,*], Qianying Hu[1,2,*], Xuan Zhao[1,2], Liguo Ma[6], Feng Jin[1], Qingming Zhang[1,3], Kenji Watanabe[4], Takashi Taniguchi[5], Jie Shan[6,7], Kin Fai Mak[6,7], Yongqing Li[1,2], Yang Xu[1,2,†]

[1]Beijing National Laboratory for Condensed Matter Physics, Institute of Physics, Chinese Academy of Sciences, Beijing 100190, China

[2]School of Physical Sciences, University of Chinese Academy of Sciences, Beijing 100049, China

[3]School of Physical Science and Technology, Lanzhou University, Lanzhou, China

[4]Research Center for Functional Materials, National Institute for Materials Science, 1-1 Namiki, Tsukuba 305-0044, Japan

[5]International Center for Materials Nanoarchitectonics, National Institute for Materials Science, 1-1 Namiki, Tsukuba 305-0044, Japan

[6]School of Applied and Engineering Physics & Department of Physics, Cornell University, Ithaca 14850, New York, USA

[7]Kavli Institute at Cornell for Nanoscale Science, Ithaca 14850, New York, USA

*These authors contributed equally to this work.
†Email: yang.xu@iphy.ac.cn

**Abstract:**
**The study of exciton-polarons has offered profound insights into the many-body interactions between bosonic excitations and their immersed Fermi sea within layered heterostructures. However, little is known about the properties of exciton polarons with interlayer interactions. Here through magneto-optical reflectance contrast measurements, we experimentally investigate interlayer Fermi polarons for 2s excitons in $WSe_2$/graphene heterostructures, where the excited exciton states (2s) in the $WSe_2$ layer are dressed by free charge carriers of the adjacent graphene layer in the Landau quantization regime. First, such a system enables an optical detection of integer and fractional quantum Hall states (e.g. v=±1/3, ±2/3) of monolayer graphene. Furthermore, we observe that the 2s state evolves into two distinct branches, denoted as attractive and repulsive polarons, when graphene is doped out of the incompressible quantum Hall gaps. Our work paves the way for the understanding of the excited composite quasiparticles and Bose-Fermi mixtures.**

The investigation of the "impurity problem", which explores the properties of a particle immersed in a quantum medium, has a profound impact in modern physics. It arises in diverse systems spanning a broad range of energy scales, including solid-state materials[1–3], ultracold atomic gases[4–6], and neutron stars[7,8]. The study focuses on elucidating the influence of the environment on a quantum object and examining how many-body systems with interactions can be comprehensively described through the utilization of "quasiparticles", which differ from the non-interacting particles by exhibiting modified static or dynamical properties. This problem was first initiated by Lev Landau in 1933 and later Solomon Pekar[9] introduced the concept of "polaron" to describe a quasiparticle formed by an excess charge carrier dressed by a cloud of virtual phonons, collective excitations in a crystal lattice. Other notable examples include helium-3 impurities in a bosonic helium-4 bath[10] and the Kondo effect caused by localized magnetic impurities in metals[11]. In recent years, the mobile quantum impurity embedded in a Fermi gas provides a paradigmatic realization of Landau's quasiparticle concept and is termed the Fermi polaron.

Lately, the Fermi-polaron model has been utilized to analyze the absorption and emission spectra of doped monolayers of two-dimensional (2D) transition metal dichalcogenides (TMDCs)[12–15]. Specifically, when an exciton (a bound electron-hole pair) is optically excited in these materials, it acts as an impurity immersed in a fermionic medium of charge carriers (electrons or holes)[12,16–18]. Their dynamical interactions lead to the emergence of two peaks in the optical response. These peaks, known as the "repulsive polaron" (RP) and the "attractive polaron" (AP), correspond to the energies that evolve into the bare exciton and a lower-energy trion (a charge-exciton bound state), respectively, as the doping concentration decreases. Gaining an understanding of how the exciton is dressed by the Fermi sea yields significant insights into the properties of the medium itself, as recently demonstrated in the TMDC monolayers and moiré superlattices[14,19–23]. However, existing studies have primarily focused on intra-layer interactions, with a notable lack of investigations exploring spatially separated interlayer interactions on the property of exciton polarons.

In this study, we report observation of interlayer Fermi polarons for the exciton excited states in $WSe_2$/graphene heterostructures using the reflection contrast measurements. The interlayer interaction is exaggerated for the $WSe_2$ excited exciton states (also referred to as Rydberg excitons) with extended electron-hole separations, when flat Landau levels (LLs) develop in graphene under out-of-plane magnetic fields. Experimentally, we observe the 2s state evolves into two distinct branches as graphene is doped out of the incompressible quantum Hall gaps.

This phenomenon strongly resembles the exciton-polaron formation in single-layer TMDCs. The energy splitting between the two branches (interlayer AP and RP, respectively) approximately increases linearly at elevated doping densities. Additionally, such interlayer Fermi polarons can serve as a microscopic and local probe for exploring both excitations and collective properties of the graphene, enabling the optical detection of the fractional quantum Hall states at v=±1/3, ±2/3. We also find that graphene exhibits the same screening property, i.e., 2D polarizability, in the gaps of the empty or filled zeroth LL. Our experimental findings allow for a comprehensive exploration of the nonlocal polaron formation within heterojunctions and a further examination of the physics in systems characterized by flat bands.

The optical detection of the graphene dielectric change is enabled by placing it adjacent to a monolayer $WSe_2$ sensor that has strong light-matter interactions due to excitonic states[24–31]. The stack of graphene/$WSe_2$ is sandwiched in hBN/graphite top and back dielectrics/gates. As depicted in Fig. 1a, the electric field lines between the electron-hole pair of an exciton extend substantially out of the $WSe_2$ monolayer. Especially for the loosely bound exciton exited states whose radii are much larger than the thickness of the atomic layers, their resonance energies and oscillator strengths are thus sensitive to the dielectric environment. The optical measurements are performed in a close-cycle cryostat at different out-of-plane magnetic fields $B$ and a base temperature $T$ of ~1.7 K (see Supplementary Information for further details). A collimated and circularly polarized white light beam with a diameter about 1 μm is focused on the sample for the reflection contrast $\Delta R/R_0 = (R-R_0)/R_0$ spectroscopy measurements.

When an out-of-plane magnetic field is applied, discrete LLs are developed in both the graphene and $WSe_2$ layers. Fig. 1c presents the density-dependent reflection contrast ($\Delta R/R_0$) spectrum of sample D obtained at $B$= 9 T. Owing to the band alignment of these two layers as illustrated in Fig. 1b, only the graphene can be effectively tuned in the whole gating range. The 1s exciton in $WSe_2$ dominates the optical spectrum around 1.71 eV and barely exhibits any doping dependence due to the combined effect of band gap renormalization and binding energy modification[26,27]. However, drastic changes at higher energies emerge as compared with its zero-field counterpart (see SI). First, more optical transitions are discerned at higher energies, corresponding to the exciton excited states $N$s (where $N = 2, 3, 4, 5, \ldots$). As the binding energies of the exciton excited states become negligible (either when graphene is highly doped or $N$ is large), the exciton dispersion with $B$ evolves into optically allowed inter-band transitions[32] (illustrated by the vertical blue arrows in Fig. 1b) between LLs formed in the valence and conduction bands. The energy separation approaches the exciton's cyclotron energy $\hbar e B / m_r$ (with $\hbar$, $e$, and $m_r$ being the reduced Planck's constant, elementary electron charge, and exciton's reduced mass, respectively). The $m_r$ is extracted to be $0.16 m_e$, consistent with previous studies[27,32,33]. Up to 12 excitonic states (only 2s-8s are labeled) can be identified in Fig.1c, indicating high quality of our samples.

Secondly, we have observed comb-like features along the exciton excited states (most pronounced for the 2s state) with tuning the charge density. They represent periodically enhanced oscillator strength due to reduced screening when the Fermi level is tuned into incompressible quantum Hall gaps of the graphene[34,35]. We know that graphene has fourfold spin-valley degeneracy with primary quantum Hall states at filling factors $\nu=\pm 2$, $\pm 6$, $\pm 10$, etc (indicated by white arrows in Fig. 1c). Under sufficiently high magnetic fields, the kinetic energies are quenched while the role of interparticle Coulomb interactions becomes crucial. Within the gate-accessible range, all other integer states (filling factor $\nu=0$, $\pm 1$, $\pm 3$, etc.) are discernable, as results of interaction driven lifting of the four-fold degeneracy and forming quantum Hall isospin ferromagnetic states (though dominant aligning field might be the Zeeman energy for the high-energy LLs)[34]. The sensitivity of exciton dielectric sensor enables the observation of LLs in the whole gating range up to LL filling factor $\nu=14$, suggesting the effectiveness of this optical detection method[27,28,35].

Thirdly, the energies of the Rydberg exciton states exhibit intriguing dependences with the carrier density $n$ or equivalently the filling factor $\nu$. For integer values of $\nu$ and $|\nu|\leq 2$, i.e., when the Fermi level aligns inside the gaps throughout the zeroth LL of graphene, the exciton energy levels remain unchanged. We will come to this point later in Fig. 4. Whereas for larger values of $|\nu|$, the excited excitons feature monotonic redshifts, mainly due to the enhanced dielectric screening effect of highly doped graphene and suppression of the quasiparticle bandgap[27].

In Fig. 1d, we provide detailed information on the density-dependent $\Delta R/R_0$ near the 2s states, as shown in the upper panel (zoomed-in view of Fig. 1c), and compare it with the capacitance $C$ measurement (lower panel) conducted on the same device. The capacitance result is consistent with the literature reported for monolayer graphene[36]. The middle panel in Fig. 1d plots the averaged value of the reflection contrast for the 2s state between the two red dashed curves in the upper panel. Again, the intensity modulation demonstrates a periodic enhancement of the oscillator strength inside LL gaps of graphene. The total capacitance $C=C_gC_q/(C_g+C_q)$, where $C_g$ and $C_q = e^2\frac{dn}{d\mu}$ are the geometric capacitance and quantum capacitance, respectively[37]. The $C_q$ is a measure of thermodynamic density of states (DOS) or the electronic compressibility[38,39]. The $\Delta R/R_0$ and $C$ measurements coincide for the gap position as indicated by the vertical dashed lines (only drawn for a few representative states) in Fig. 1d, where the oscillator strength maxima (dips in $\Delta R/R_0$) correspond to the dips in $C$, indicating the locations of the DOS minima. However, we note that the strength in $\Delta R/R_0$ does not have a one-to-one correspondence on the dip value in $C$ nor the gap size value which can be obtained by integrating $\frac{1}{C_q} \propto \frac{d\mu}{dn}$ over $dn$. Besides, we observe that the dip features in $\Delta R/R_0$ is considerably wider than that in $C$ (especially true for the interaction-driven gaps at ν=0, ±1, ±3, etc), likely due to the preservation of the excited excitonic states when dressed by finite charge carriers inside the LLs.

We then focus on the interlayer Fermi polaron feature observed near the charge neutral region for the excited exciton states (Fig. 2, measured for sample A). Within the zeroth LL gap at the charge neutrality (ν = 0), higher excitonic states (2s, 3s, 4s, 5s, etc) are observed (Fig. 2a). However, when carrier density is doped away from ν = 0, the higher excitonic states with $N \geq 3$ rapidly vanish. As the carrier density increases, the 2s state exhibits a blue shift accompanied by the emergence of a lower-energy resonance state. This is clearly depicted in Fig. 2b by the orange linecuts from the two dashed boxes in Fig. 2a. As the carrier density increases, notably, the lower energy branch features an enhancement in strength at ν = ±1/3, ±2/3, providing evidence for the presence of incompressible fractional quantum Hall states in this high-quality sample. Another phenomenon that occurs with increasing the carrier density is a gradual weakening of the strength of the high-energy branch and a corresponding enhancement in the strength of the low-energy branch, until the carrier density reaches about |ν|=1/2. The observed energy splitting behavior strongly resembles that of exciton polarons formed during the doping of TMDC monolayers. In that case, the relative transfer of spectral weight from AP to RP as the Fermi energy of monolayer TMDCs increases is due to an increase in the exciton component of the AP branch[14,15,17,18]. We hence conclude that in our system, the presence of excess electron doping in the LL of graphene causes the excitonic 2s state in WSe$_2$ to split into two distinct components: high-energy blue-shifted RPs and low-energy APs. As a comparison to the intralayer exciton polarons, we provide an illustration of the attractive and repulsive interactions for the interlayer Fermi polarons for our system in Fig. 2c. The AP is the low-energy favorable state, while the RP is the higher-energy metastable excitation. We note that the larger spatial extension (~7 nm for the 2s) of the excited exciton states, along with the flattened dispersion of LLs, is crucial for such observations, as such effects has not been observed for the ground state exciton (1s state) nor at zero magnetic fields. Subsequently, with further increasing the charge density, there is a reversed process of transferring the spectral weight, which continues till entering the next LL gaps (|ν|=1). This is related to the charge carriers within the Landau level changing from electron-like to hole-like across the half filling.

To gain further insights about the interlayer Fermi polarons in our system, we analyze the spectral weights, linewidths, and energies of both the APs and RPs near ν = 0 in Fig. 3 (see details in SI). First, the density dependence of the AP spectral weight shows an opposite trend as that of the RP, as depicted in the upper panel of Fig. 3a. The evolution of spectral weights represents the proportion of the two polaron branches in the scattering processes between excitons and carriers in the Fermi sea, indicating a transfer of oscillator strength from the metastable RP branch to the AP branch with increasing charge doping. Meanwhile, the linewidth (Γ) of the AP remains relatively small (1~2 meV) in the doping range, corresponding to the intrinsic broadening limit. This is in direct contrast to ~10 meV broadening by autoionization for the intralayer 2s AP observed in monolayer WSe$_2$[41], indicating absence of such scattering processes in our system. On the other hand, the RP features a rapid increase of linewidth from ~2 meV to ~5 meV with increasing density. Despite the interlayer and enlarged orbital nature (associated with the 2s states), such linewidth evolution is similar to that of the intralayer exciton polarons found in the monolayer TMDC. The sharp resonance of AP coincides with the intrinsic broadening of excitons due to the bound nature. In contrast, the RP consists of a continuum of eigenstates with closely spaced eigenvalues, resulting in a broadened linewidth at higher doping levels[18,42].

We then investigate the polaronic splitting of the 2s excitonic state, which provides insights into the strength of the interaction between the exciton and the excess charges in the Fermi sea. Fig. 3b illustrates the density dependences of the energy splitting $\Delta E_{RP-AP}$ between the RP and the AP near $\nu = 0$ under different vertical magnetic fields. As can be seen, $\Delta E_{RP-AP}$ increases roughly linearly with the density either on the electron or the hole doped sides. The data at a given field can be phenomenologically fitted using the following equation: $\Delta E_{RP-AP} = E_T + \beta \frac{\hbar^2}{m_e} n$ (with $\frac{\hbar^2}{m_e} \approx 0.75 \times 10^{-12}$ meV·cm$^2$) and the fitting results are depicted by the solid lines. Here, $E_T$ represents the binding energy of the AP in the limit of vanishing charge density, referred to as a trion (exciton bound to one charge). The $\beta$ is a dimensionless prefactor reflecting the energy splitting rate with density change. We plot the extracted $E_T$ for this sample (labeled as sample A) and another two samples (B and C, respectively) in Fig. 3c, where its evolution with the magnetic field is guided by the dashed curve. First, at zero magnetic field limit, there is no discernible energy splitting of the 2s state in our system (see SI), indicating $E_T = 0$, which is likely related to the vanishing effective mass of Dirac fermions in graphene. The formation of LL flatbands under magnetic fields facilitates a finite trion binding energy $E_T$ for the WSe$_2$ 2s exciton with the LL charge carriers in graphene. Within the experimental magnetic field range (6 ~ 12 T), the magnitude of $E_T$ grows from 4 to 8 meV, being much smaller than $E_T$ for intralayer exciton-polarons (14 ~20 meV, observed for the 1s and 2s states) in monolayer TMDCs[14,41]. The saturation trend at higher magnetic fields may be related to the magnetic length $l_B = \sqrt{\frac{\hbar}{eB}} = \frac{25.7}{\sqrt{B}}$ nm (specifying the interparticle distance in a LL and the interparticle Coulomb energy $\propto 1/l_B$) becoming comparable to the radius of the 2s exciton (~7 nm)[32].

For the intralayer Fermi polarons in monolayer TMDCs, the splitting $\Delta E_{RP-AP}$ is expected to grow linearly with the Fermi energy $E_F$ of the Fermi sea as $\frac{m^*}{m_T}E_F = \beta\frac{\hbar^2}{m_e}n$ at low densities, with $m^*$ being the electron effective mass and $m_T$ being the reduced exciton-electron mass[16]. For the 1s state, considering similar electron and hole masses that gives rise $m_T \approx \frac{2}{3}m^*$, $\beta$ is obtained about 11.8 (WSe$_2$ with $m^* \approx 0.4m_e$) and 5.9 (MoSe$_2$ with $m^* \approx 0.8m_e$), in good agreement with experiments[12,14–16,41]. In the case of the 2s state, the enhanced interaction between excitons and the Fermi sea, due to the larger spatial extension of the exciton excited states, leads to a steeper slope with $\beta = 16\sim19$ for monolayer WSe$_2$ [14,41]. Our study shows the absolute value of $\beta$ for the 2s interlayer Fermi polaron ranging from 19 to 55 (inset of Fig.3c). It in general shows a decreasing trend with increasing the magnetic field and exhibits large fluctuations among different samples. We hypothesize that the finite value of $\beta$ is likely related to the LL broadening, which could also explain the variation due to the difference in sample quality. However, future theoretical studies are needed for a better understanding of the interlayer Fermi polarons for the excited exciton states in the quantum Hall regime.

Lastly, we briefly discuss about another intriguing observation for the excitonic states when the Fermi level is within the zeroth LL gaps at ν = 0, ±1, ±2. In such scenarios, there are no free charge carriers present in graphene and the interlayer interaction is primarily influenced by a static screening effect. Under the same magnetic field (e.g. $B$=9 T), the energies of excited exciton states with the same quantum number ($N$) are identical at these filling factors (ν = 0, ±1, ±2) as illustrated by the vertical dashed lines in Fig. 4a. We illustrate the robustness of this behavior at a few magnetic fields from 5 T to 9 T in Fig. 4b. The excited $N$s states continuously spreads out with increasing the magnetic field (as combined effects of Zeeman energy, diamagnetic shift, and cyclotron energy[32]), whereas their energies remain the same for ν = 0, ±1, ±2 at the same magnetic field (see collapsed curves in Fig.4b). It indicates excitons in WSe$_2$ experience identical screening properties through different zeroth LL gaps (at ν = 0, ±1, ±2), despite these gaps varying in size and originating from different mechanism (single-particle for ν =±2 and interaction-driven for ν =0, ±1). This phenomenon is likely attributed to the consistent 2D polarizability within the gaps of the splitted zeroth LLs, combined with the particle-hole symmetry in graphene.

In summary, we have experimentally investigated the excited exciton states in WSe$_2$ dressed by free charge carriers confined within LLs of adjacent graphene. This interaction leads to the formation of a novel type of exciton complexes, termed as interlayer Fermi polarons for the excited exciton states. When compared to their intralayer counterparts, such interlayer Fermi polarons allow us to identify several distinct features regarding the degeneracy-lifted LLs in graphene and the spatially-separated polaronic interactions. The dielectric screening experienced by the excited excitons also enables optical detection of the many-body quantum states (such as the interaction-driven quantum Hall ferromagnets and fractional quantum Hall states) and polarizability of the graphene layer. These findings may shed new light on the many-body quantum states and understanding unconventional quantum Hall physics.

**Notes**

The authors declare no competing financial interest.

**Acknowledgements**

We thank F. Wu and A. H. MacDonald for helpful discussions. This work was supported by the National Key Research and Development Program of China (Grant Nos. 2021YFA1401300, 2022YFA1403403, and 2022YFA1402704), the National Natural Science Foundation of China (Grant No. 12174439), and the Innovation Program for Quantum Science and Technology (Grant No. 2021ZD0302400). The work at Cornell University was supported by the U.S. Department of Energy, Office of Science, Basic Energy Sciences, under Awards No. DE-SC0019481. The growth of hBN crystals was supported by the Elemental Strategy Initiative of MEXT, Japan, and CREST (JPMJCR15F3), JST.

**Figures:**

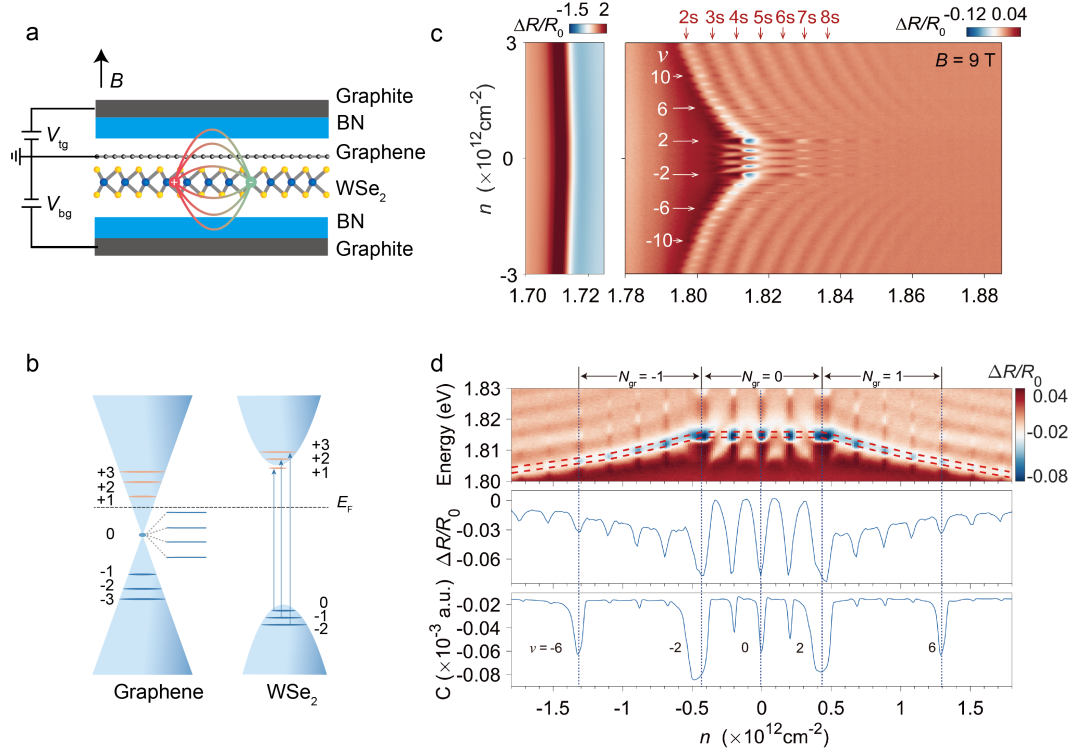

**Figure 1. Dielectric sensing of the quantum Hall states in graphene with excited excitons. a.** Schematic of the device structure. Coulomb interactions in WSe₂ are sensitive to the dielectric change of the adjacent graphene layer. The arrow represents the out-of-plane direction of the applied magnetic field. **b.** Schematic showing the band alignment and Landau level (LL) formation (horizontal lines) in graphene and WSe₂. Vertical arrows indicate interband transitions in WSe₂. **c.** Density-dependent reflection contrast ($\Delta R/R_0$) spectra measured with circularly polarized light at $B$= 9 T for sample D. Red arrows denote the WSe₂ exciton excited states ($N$s states, with $N$=2, 3, 4, …), which merge into the interband LL transitions (-1↔+2, -2↔+3, -3↔+4, …, as depicted for only one valley) at vanishing binding energies. The periodical enhancements in the oscillator strength of the excitonic states correspond to the incompressible gaps of the integer quantum Hall states in graphene, where the four-fold degeneracy of each LL is fully lifted (only depicted for the zeroth LL in **b**). White arrows indicate fully filled LLs at filling factor v= ±2, ±6, ±10. **d.** Upper panel is a zoomed-in plot of **c**, whereas the middle panel and lower panel depict the -$\Delta R/R_0$ (averaged between the two red

dashed curves) and capacitance (*C*) measured for the same device as functions of the carrier density, respectively. Vertical dashed lines indicate filling factors at ν = 0, ±2, ±6. The $N_{gr}$ denotes the graphene LL orbital index.

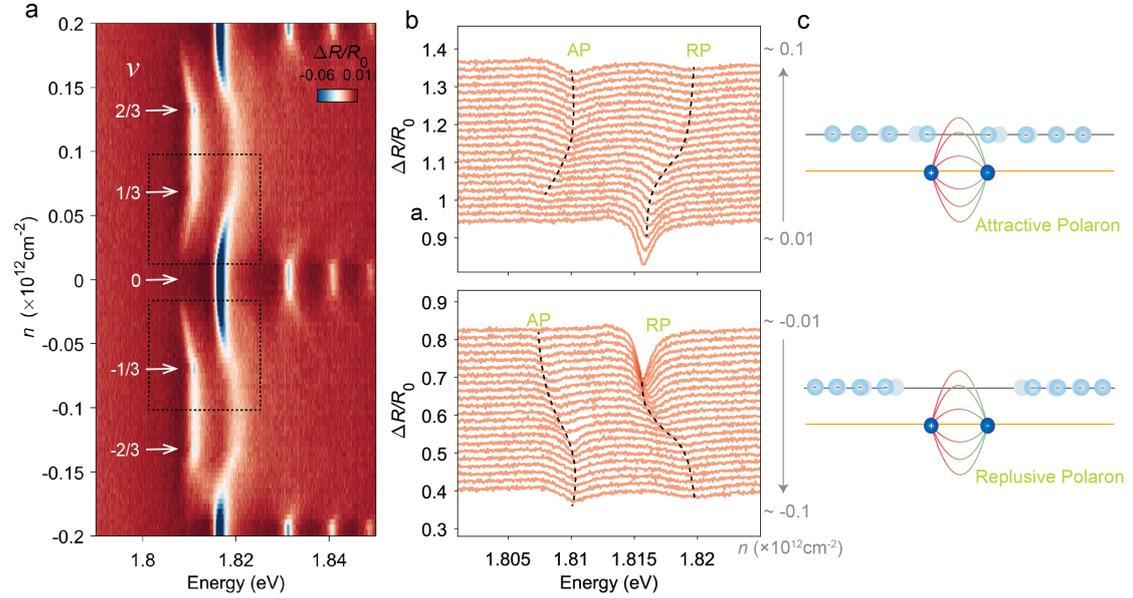

**Figure 2. Interlayer attractive/repulsive polarons (AP/RP) of the 2s states and fractional quantum Hall states near ν=0 at *B*=9 T in sample A. a.** Density-dependent reflection contrast ($\Delta R/R_0$) spectra for the carrier density ranging from -0.2×10$^{12}$ to 0.2×10$^{12}$ cm$^{-2}$ for sample A. The 2s states split into two branches, i.e. attractive polarons (APs) and repulsive polarons (RPs) when doped out of the incompressible gap at ν=0. The AP branches show enhancements of intensity at fractional filling factors of ν=±1/3, ±2/3, indicating the emergence of fractional quantum Hall states and higher quality of this sample. **b.** Selective linecuts (vertically shifted for clarity) of **a** with the upper and lower panels corresponding to the filling factor range 0 < ν < 1/2 and -1/2 < ν < 0 (inside the two dashed boxes in **a**), respectively. The dashed curves indicate the evolution of the APs and RPs with the density. **c.** Schematic diagrams of the interlayer attractive/repulsive polarons.

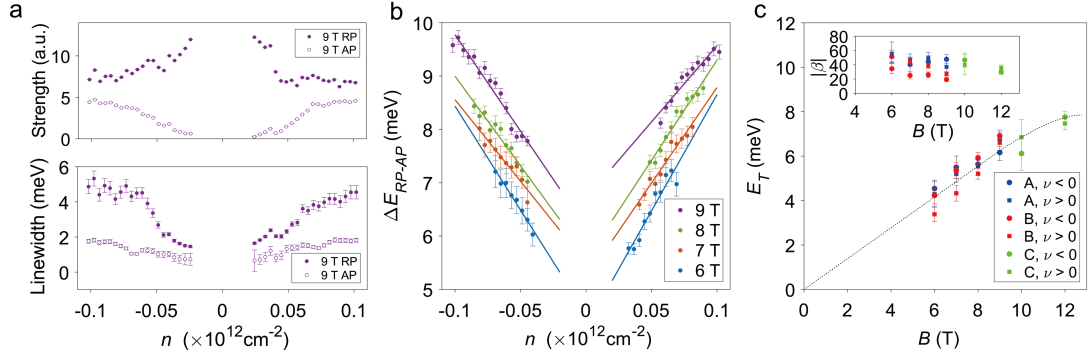

**Figure 3. Evolution of AP and RP with the carrier density and magnetic field. a.** Spectral weight (Upper panel), linewidths (lower panel) of the APs and RPs as functions of the carrier density $n$ near $\nu=0$ at $B=9$ T. As the carrier density increases, the spectral weight of RPs decreases while that of APs increases. Meanwhile, the linewidth ($\Gamma$) of RPs increases in a faster fashion than that of the APs. **b.** The energy splitting between the APs and RPs as a function of $n$ at $B=6$, 7, 8, 9 T. The solid lines represent linear fits to the experimental data, characterized by the equation $\Delta E_{RP-AP} = E_T + \beta \frac{\hbar^2}{m_e} n$. At vanishing densities, $E_T$ represents an interlayer trion binding energy. **c.** The evolution of $E_T$ and $|\beta|$ (shown in the inset) with the magnetic field summarized for different samples. The dashed curve denotes a trend of saturation with increasing the magnetic field. Moreover, the slope $|\beta|$ generally decreases with increasing the magnetic field within the same sample and exhibits larger variations among different samples.

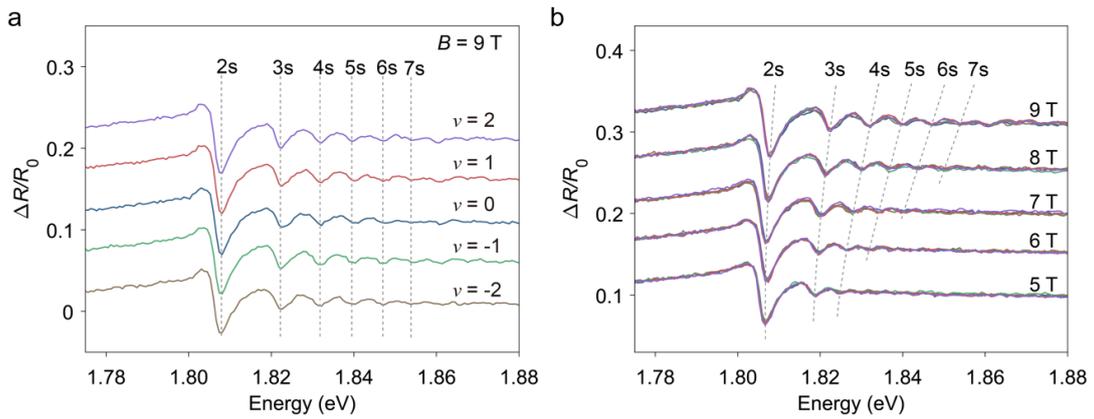

**Figure 4. The identical dielectric screening effect at different zeroth LL gaps of graphene ($N_{gr}=0$). a.** Representative reflection contrast ($\Delta R/R_0$) spectra at $B=9$ T at $\nu=0, \pm1, \pm2$. The vertical dashed lines denote the same energies for the $N$s excitons (ranging from 2s to 7s), indicating the same dielectric screening effect at $\nu=0, \pm1, \pm2$. **b.** Representative linecuts at $\nu$

= 0, ±1, ±2 (collapsed for each field) under different magnetic fields. The dashed curves trace the magnetic field induced blueshifts of the excited exciton states.

**Supporting Information for "Interlayer Fermi polarons of excited exciton states in quantizing magnetic fields"**

**Sample preparation**

We mechanically exfoliate the monolayer WSe$_2$, graphene, hBN and graphite from bulk crystals and fabricate van der Waals heterostructures using a modified layer-by-layer dry-transfer method[1]. The heterostructures are released onto silicon substrates with a 285-300 nm oxide layer and prepatterned Cr/Au electrodes (see a typical device image in Fig. S1 and structure in Fig.S2a). We have shown experimental results obtained from four different samples in the main text (sample A-D). Despite slight differences in the sample quality, the general gate-induced optical behaviors are the similar for all the four samples.

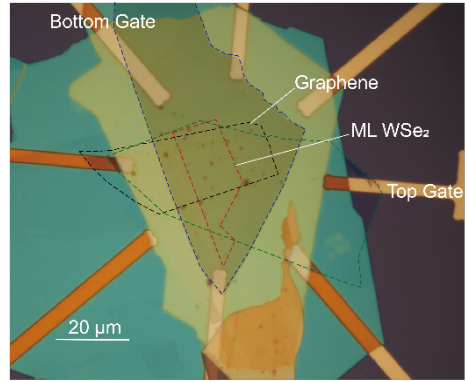

**Fig S1. Optical image of a typical device (sample B).**

**Optical measurements**

Optical measurements are mainly performed with the devices loaded inside an attoDRY2100 (Attocube) cryostat, allowing the device to be cooled to 1.7 K and the application of out-of-plane magnetic fields up to 9 T. A halogen lamp is used as a white light source, whose output is collected by a single-mode fiber and then collimated by a 10× objective. Magneto-optical measurements are performed with a circularly polarized light, which is generated by a combination of a linear polarizer and an achromatic quarter-wave plate. The beam is focused onto the sample by an objective with a numerical aperture of 0.82. The reflected light is collected by the same objective and the spectrum is obtained from a spectrometer. The acquired reflectance signals are presented in terms of reflection contrast, defined as $\Delta R/R_0 = (R - R_0)/R_0$, where $R$, $R_0$ denote reflected light spectrum from the sample and the bare substrate, respectively. Fig. 1c is measured with a grating of 600 g/mm and a coarse gate scanning for sample D. Fig. 2 is measured with a grating of 1800 g/mm and a fine gate scanning for sample A, which is a higher quality sample showing sharper excitonic resonances. Additional optical measurements up to 12 T for sample C is performed with similar setups in a PPMS DynaCool (Quantum Design) cryostat.

**Reflection contrast at *B* = 0**

The density-dependent reflection contrast ($\Delta R/R_0$) spectrum of sample B obtained at *B*= 0 T is depicted in Fig S2c. Owing to the band alignment of these two layers as illustrated in Fig. 1b, only the graphene can be effectively doped (with charge carrier density *n*) and the WSe$_2$ remains charge neutral in the whole gate tuning range. At the graphene Dirac point where *n*=0 (linecut shown in Fig. S2d), the spectrum is dominated by the 1s exciton (near 1.71 eV) of WSe$_2$, and the discernible 2s and 3s excitons (near 1.80 eV) evolve into the continuum at higher photon energies. Upon increasing either electron or hole doping, the modification to the dielectric function of graphene renders a renormalized quasiparticle bandgap $E_g$ and reduced exciton binding energies $E_b^{(Ns)}$ (for the *N*s exciton, principle quantum number *N*=1, 2, 3, etc.) of WSe$_2$. For the 1s exciton, these two contributions nearly cancel each other, resulting in a nearly doping-independent resonance energy and oscillator strength as shown in the left panel of Fig. S2c[2–4]. Whereas for the exciton excited states with much larger radii, the long-range Coulomb interaction easily gets screened by the excess charge carriers from graphene. The 2s/3s states merge into the band edge transition, whose energy decreases linearly with the graphene density of state ($DOS \propto \sqrt{|n|}$) and exhibits a mirror symmetry relative to the Dirac point[4], reflecting the electron-hole symmetry of the massless Dirac fermions.

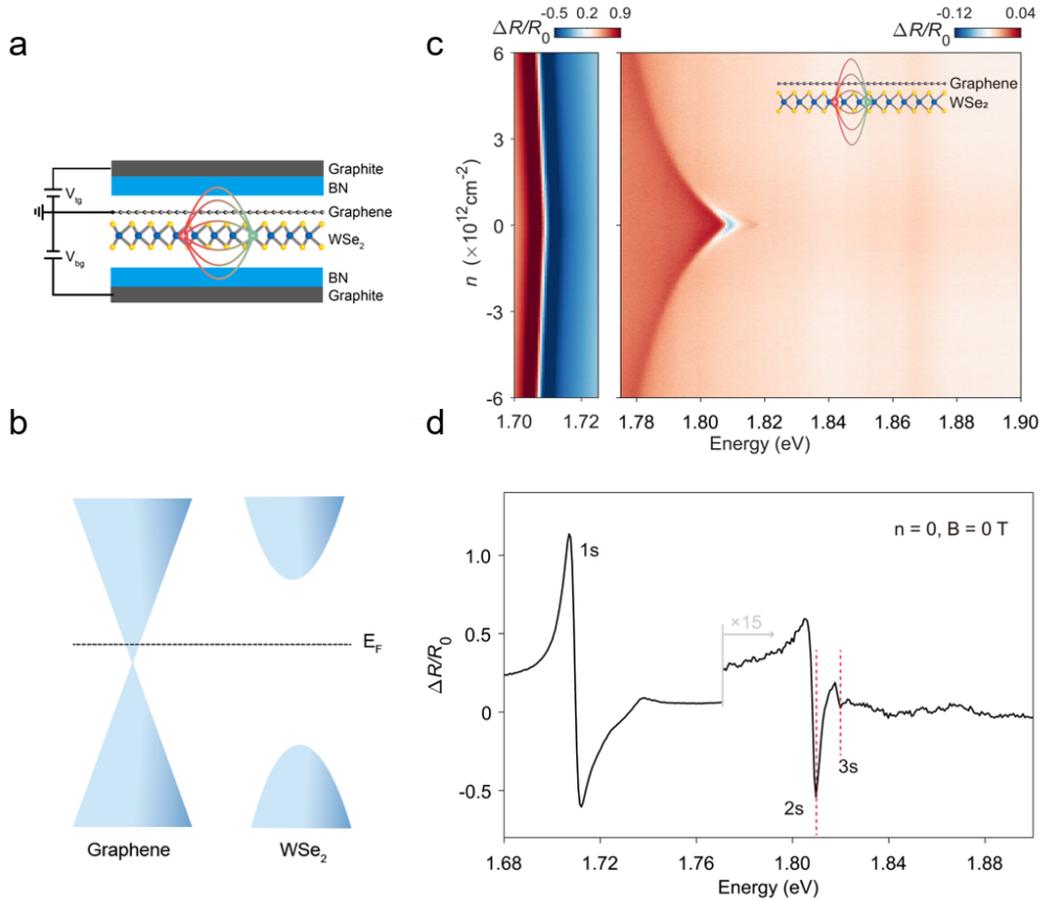

**Fig S2. Reflection contrast of sample D at *B* = 0. a.** Schematic of the device structure. **b.**

The band alignment between graphene (left) and monolayer WSe₂ (right) illustrates the Dirac point of graphene deeply buried within the band gap of WSe₂. **c.** Density-dependent reflection contrast ($\Delta R/R_0$) spectra measured at $B = 0$. **d.** The linecut of **c** at $n = 0$.

## Capacitance measurement

Typical quantum capacitance measurement for small devices is applied to sample D (results shown in the lowest panel of Fig.1d) for direct comparison with the reflection contrast spectra. A commercial high electron mobility transistor (model FHX35X) is used to create a low-temperature capacitance bridge-on-a-chip[5–7]. This configuration effectively isolates the device capacitance from the parasitic capacitance induced by cryogenic cabling. The differential capacitance, labeled as $C$, is determined by applying an AC voltage (~10 mV amplitude and 3–5 kHz frequency) between the gate and graphene. Charges were collected from the gate through the transistor using lock-in techniques.

## Fitting of the reflection contrast

The measured reflectance contrast ($\Delta R/R_0$) spectra mix the real and imaginary parts of the dielectric function of WSe₂/Graphene through the optical interference in the BN/graphite /BN/WSe₂/Graphene/BN/graphite/SiO₂/Si heterostructure. We can extract the imaginary part of dielectric function by solving the optical problem in our stacked material system with the transfer matrix method in linear optics[8]. The thickness of the SiO₂ layer is 300 nm. The thickness of the top and bottom hBN layers are 34 nm and 32 nm and the thickness of the top and bottom graphite layers are 3 nm and 5 nm, acquired by atomic force microscopy (AFM) for sample B. The refractive index of hBN is taken to be $n_{BN}$~2.1 for the frequency range of our experiment[9,10].

Our method of extracting the dielectric function is similar to the Kramers–Kronig constrained variational analysis developed by A. B. Kuzmenko[11]. The energy-dependent dielectric function of the WSe₂/Graphene, denoted as $\varepsilon(E)$ is characterized by the parameterization utilizing multiple Lorentzian resonances:

$$\varepsilon(\text{E}) = \varepsilon_\infty + \sum_{j=1}^{N} \frac{f_j}{E_j^2 - E^2 - iE\Gamma_j} \ , \quad (1)$$

where $\varepsilon_\infty$ is the high-frequency dielectric constant, which represents the contribution of all oscillators at very high frequencies. The $f_j$, $E_j$, $\Gamma_j$ represent the oscillator strength, peak energy, and the damping of the $j$-th oscillator, respectively.

Subsequently, by incorporating equation (1) and considering the reflection of multiple layers, we fit the resonance peaks to extract information such as the vibrational intensity, linewidth, and energy. The energy-dependent dielectric function of the WSe₂/Graphene heterostructure, denoted as $\varepsilon(E)$, is characterized by the parameterization utilizing multiple Lorentzian resonances. Subsequently, by incorporating equation (1) and considering the reflection of multiple layers, we fit the resonance peaks to extract information such as the vibrational intensity, linewidth, and energy (see an example in Fig. S3).

For the main device discussed in the manuscript (sample A), due to the relatively thick BN layers, the shape of the reflection spectrum resembles a Lorentzian profile. Thus, we directly employ the Lorentzian equation to fit the reflection contrast, which yields results consistent with those obtained from optical transfer matrix method. Consequently, we utilize the Lorentzian line shape to fit the reflection spectrum (see Fig. S4).

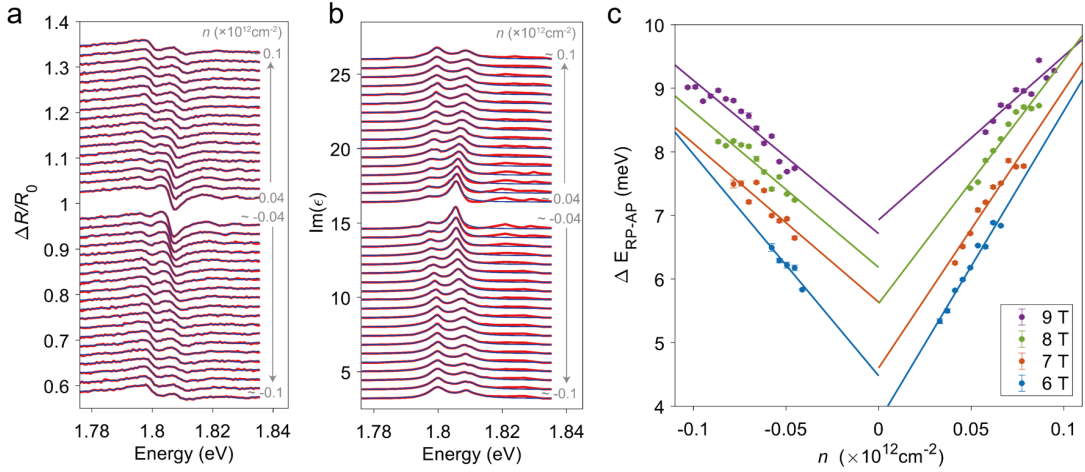

**Fig S3. Interlayer attractive/repulsive polarons of exciton excited states of sample B at $B$ = 9 T. a.** Red curves: reflection contrast at $B$ = 9 T. The upper and lower sets of lines are in the region where $0 < \nu < 1/2$ and $-1/2 < \nu < 0$, respectively. The blue curves are fits by the Kramers–Kronig constrained variational method which we can use to extract the resonance energies, oscillator strengths, and linewidths. **b.** Red curves: extracted imaginary parts of the dielectric function of monolayer WSe$_2$/graphene layer for corresponding lines in **a.** Blue curves: imaginary parts of the parameterized dielectric function used in multi-Lorentzian simulation. The 2s state splits into two branches, with two dashed lines indicating attractive polarons (APs) and repulsive polarons (RPs) at finite carrier densities. **c.** The energy splitting between the APs and RPs as a function of $n$ at $B$=6, 7, 8, 9 T. The solid lines represent linear fits to the experimental data.

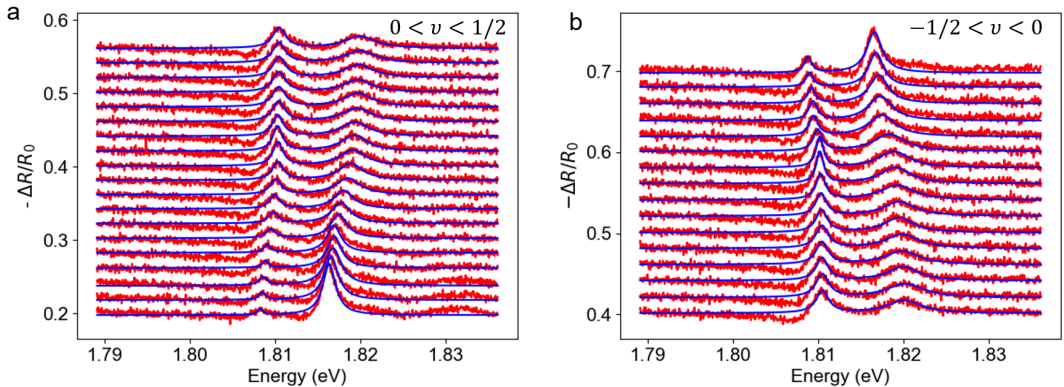

**Fig S4.** Comparison between the fitted spectra (blue curves) and the original spectra (red) for sample A.

Energy splitting of the AP and the RP at **-2 < ν < 2**

Within the range of -2 < ν < 2, the extracted energy separation $\Delta E_{RP-AP}$ varies linearly with the gate voltage.

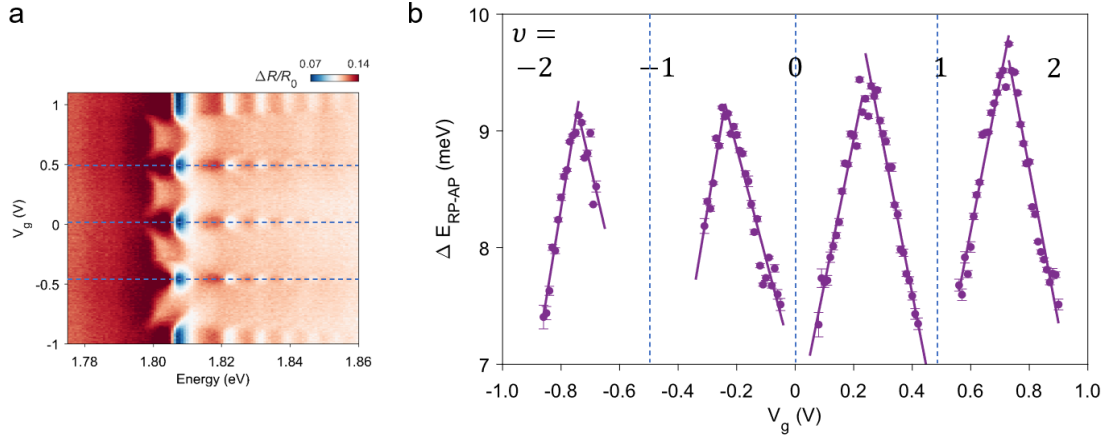

**Fig S5. Energy separation between interlayer attractive/repulsive polarons of exciton excited states at 9 T for sample B**. **a**, Gate-dependent reflection contrast spectrum of sample B at $B = 9$ T. **b**, Energy separation $\Delta E_{RP-AP}$ between attractive/repulsive polarons between ν= -2 and ν= 2.